\documentclass[twoside,11pt]{article}

\usepackage{blindtext}

\usepackage[preprint]{dmlr2e}
\usepackage{soul}
\usepackage{amsmath}
\usepackage{booktabs}
\usepackage{threeparttable}
\usepackage{color} 
\usepackage{etoolbox}
\usepackage{longtable}
\usepackage{tabularx,booktabs, array}
\usepackage{bbm}
\newcolumntype{Y}{>{\raggedright\arraybackslash}X}
\newcolumntype{P}[1]{>{\raggedright\arraybackslash}p{#1}}

\makeatletter
\patchcmd{\@maketitle}
  {\\ \\ {\bf Reviewed on OpenReview:} \openreview}
  {}{}{}
\makeatother

\usepackage{lastpage}

\ShortHeadings{Measurement Error Labeling}{Chew et al.}
\firstpageno{1}

\begin{document}

\title{From Ground Truth to Measurement: \\ A Statistical Framework for Human Labeling}

\author{\name Robert Chew\textsuperscript{*} \email rchew@rti.org
       \AND
       \name Stephanie Eckman\textsuperscript{\dag} \email steph@umd.edu
       \AND
       \name Christoph Kern\textsuperscript{\ddag\S}  \email christoph.kern@lmu.de
       \AND
       \name Frauke Kreuter\textsuperscript{\dag\ddag\S} \email fkreuter@umd.edu \\
       \AND
       \addr \textsuperscript{*}Department of AI, Data, and Product Engineering, RTI International \\
       \addr \textsuperscript{\dag}Social Data Science Center, University of Maryland \\
       \addr \textsuperscript{\ddag}Department of Statistics, LMU Munich \\
       \addr \textsuperscript{\S}Munich Center for Machine Learning (MCML)
    }


\maketitle

\begin{abstract}%
Supervised machine learning assumes that labeled data provide accurate measurements of the concepts models are meant to learn. Yet in practice, human labeling introduces systematic variation arising from ambiguous items, divergent interpretations, and simple mistakes. Machine learning research commonly treats all disagreement as noise, which obscures these distinctions and limits our understanding of what models actually learn. This paper reframes annotation as a measurement process and introduces a statistical framework for decomposing labeling outcomes into interpretable sources of variation: instance difficulty, annotator bias, situational noise, and relational alignment. The framework extends classical measurement-error models to accommodate both shared and individualized notions of truth, reflecting traditional and human label variation interpretations of error, and provides a diagnostic for assessing which regime better characterizes a given task. Applying the proposed model to a multi-annotator natural language inference dataset, we find empirical evidence for all four theorized components and demonstrate the effectiveness of our approach. We conclude with implications for data-centric machine learning and outline how this approach can guide the development of a more systematic science of labeling.
\end{abstract}


\section{Introduction}

Machine learning models are only as good as the data they learn from. In supervised learning, that data depends on human annotators who assign labels to training and evaluation examples. A single mislabeled instance can propagate downstream into model errors \citep{frenay2013classification, song2022learning, northcutt2021confident}, fairness disparities \citep{sap-etal-2022-annotators, paulus2020predictably}, and misleading benchmarks \citep{northcutt2021pervasive}. Labeling is thus not only a practical bottleneck but also a scientific one: the reliability of labeled data constrains what supervised models can learn \citep{sculley2015hidden, sambasivan2021everyone} and how we can meaningfully evaluate them \citep{paullada2021data, geiger2021garbage}.

Despite these well-documented effects, relatively little work examines the labeling process itself, such as how and why annotation errors arise, and what factors make certain instances, annotators, or contexts more error-prone. Existing approaches often treat labeling errors as random or adversarial that can be corrected post hoc rather than as a structured outcome of human judgment and cognitive effort. Understanding this structure is essential if we wish to model, mitigate, or meaningfully interpret labeling error.

The social and behavioral sciences have long confronted analogous problems under the broader concept of \emph{measurement error}. In survey research and psychometrics, measurement models formalize how observed responses deviate from latent ``true'' values through systematic and random components \citep{groves2011survey,lord2008statistical}. These frameworks provide principled tools to reason about concepts such as reliability, validity, and bias that map directly onto challenges in human labeling and evaluation for machine learning.

This paper adapts and extends the classical measurement error model to annotation for supervised learning tasks. We first recast annotation as a measurement process, in which each observed label is a noisy realization of an underlying true label. We then extend this framework to account for \emph{human label variation} \citep{plank2022problem}, recognizing that the ``true'' label may differ across annotators for subjective or value-laden tasks such as hate speech detection or sentiment analysis. The resulting model decomposes observed labeling outcomes into components attributable to the instance, the annotator, and the annotation context, offering a principled way to quantify and interpret label noise. We then empirically demonstrate how this framework isolates structured sources of annotation error on the Variation versus Error (VariErr) dataset \citep{weber-genzel-etal-2024-varierr} and provide a method to help assess which interpretive regime (global or individual ground truth) is more plausible for a given task.

More broadly, a measurement perspective complements recent calls for \emph{data-centric} machine learning \citep{jarrahi2023principles}. While most approaches to dataset quality rely on heuristic filtering or post hoc label correction, our framework provides a principled statistical basis for understanding why labeling errors occur and how they propagate through model training and evaluation. Because the framework decomposes variance into interpretable components, it can guide practical decisions about dataset construction (e.g., identifying ambiguous instances), annotator management (e.g., detecting consistent biases), and evaluation design (e.g., quantifying the reliability of benchmarks).  
Although we demonstrate the framework using a natural language inference task, the underlying logic generalizes to any supervised setting involving human annotation, from image classification to toxicity detection, where understanding the structure of disagreement is crucial for responsible model development.

\paragraph{Contributions.} 
Our contributions are both conceptual and empirical, motivated by the broader claim that a measurement error perspective offers diagnostic, theoretical, and modeling advantages for understanding label quality in supervised learning.

\begin{enumerate}
    \item \textbf{A general measurement error framework for supervised learning.}  
    We formalize annotation as a measurement process in which observed labels are noisy realizations of latent truths.  
    This provides a \emph{diagnostic basis} for analyzing error: by decomposing variance into instance-, annotator-, and residual components, the framework identifies whether labeling errors arise from ambiguous items, systematic annotator tendencies, or unstable within-person factors (Section~\ref{concept-model}).  
    These distinctions, in turn, inform how labeling tasks can be redesigned to address the underlying sources of error.

    \item \textbf{Extension to individual ground truths.}  
    We extend the classical model to tasks where multiple valid interpretations coexist, linking human label variation to long-standing notions of reliability and validity in measurement theory.  
    This extension provides \emph{theoretical clarity} about when disagreement reflects meaningful interpretive diversity rather than simple error, offering a principled bridge between psychometrics and contemporary annotation practice (Section~\ref{app:ind-truth}).

    \item \textbf{Empirical demonstration using real annotation data.}  
    Applying the framework to the Variation versus Error (VariErr) dataset, we show how instance-, annotator-, and context-level factors jointly shape labeling accuracy and disagreement.  
    The analysis reveals structured sources of variance that standard label-noise models obscure, illustrating how the framework can \emph{quantify and diagnose} distinct patterns of labeling error (Section~\ref{case-study}).

    \item \textbf{A diagnostic for identifying measurement regimes.}  
    We develop a method to evaluate whether a labeling task behaves as if it reflects a single shared ``global'' truth or structured individual variation.  
    This assessment clarifies the boundary between labeling regimes and provides \emph{modeling guidance}—indicating when aggregation, personalization, or distributional modeling of labels is most appropriate (Section~\ref{sec:diagnostic}).
\end{enumerate}

\section{Related Works}

\paragraph{Label Noise.} 
A substantial body of work in machine learning examines the effects of label noise and methods for mitigating it. This literature typically treats mislabeling as a statistical nuisance that corrupts the true signal in supervised learning. Classical studies show that even modest rates of incorrect labels can bias parameter estimates, reduce predictive accuracy, and degrade generalization performance \citep{frenay2013classification, song2022learning}. In response, recent approaches in deep learning propose algorithmic strategies to detect or correct mislabeled instances, including sample reweighting, noise-robust loss functions, and confidence-based label correction \citep{northcutt2021confident, patrini2017making, han2018co}. While these methods improve model robustness, they largely abstract away from the human and cognitive processes that generate label noise in the first place. Moreover, they often rely on simplifying assumptions, such as random or class-conditional label flips, because the real-world mechanisms that produce annotation errors remain poorly understood \citep{frenay2013classification}. Our work complements this line of research by shifting attention from algorithmic mitigation to the \emph{source} of label noise itself, treating labeling as a structured measurement process shaped by instance characteristics, annotator traits, and contextual factors.

\paragraph{Crowdsourcing and Annotation Models.}
The crowdsourcing literature provides a parallel line of work that explicitly models disagreement among annotators rather than assuming a single ground truth. Seminal probabilistic models such as Dawid and Skene's \citeyearpar{dawid1979maximum} expectation–maximization framework estimate latent true labels jointly with annotator reliability parameters, effectively distinguishing systematic annotator bias from random error. Subsequent extensions, such as the Generative model of Labels, Abilities, and Difficulties (GLAD) \citep{whitehill2009whose}, introduce item difficulty and annotator expertise as interacting latent variables, formalizing how harder instances and less reliable annotators jointly contribute to labeling noise. More recent Bayesian and neural variants further refine these ideas for large-scale crowdsourced datasets \citep{raykar2010learning, chu2021learning}. Our work complements this literature by drawing on its insight that annotation errors are structured and predictable, but reframes the problem through the lens of measurement theory. Whereas crowdsourcing models primarily focus on recovering a consensus or estimating annotator accuracy, our approach decomposes annotation variability into distinct, interpretable components, providing a complementary perspective on the factors that give rise to disagreement.

\paragraph{Annotation Disagreement and Human Label Variation.} A growing line of research in natural language processing and human-centered machine learning challenges the assumption that there exists a single, objective “ground truth” label for every instance. Instead, it emphasizes that annotators bring diverse perspectives, experiences, and values to labeling tasks, leading to systematic variation in their judgments \citep{fleisig2024perspectivist, prabhakaran2021perturbing, sorensen2024plural}. Such human label variation (HLV) is particularly salient in socially and linguistically subjective domains such as toxicity detection, hate speech, or sentiment analysis, where disagreement often reflects pluralism in meaning rather than annotator error \citep{aroyo2015truth, plank2022problem}. Recent work has therefore argued for modeling and preserving disagreement rather than collapsing it into a single consensus label \citep{uma2021learning, davani2022dealing, orlikowski2023ecological}. Our approach builds on these insights by offering a measurement-theoretic perspective that helps distinguish between disagreement arising from interpretive variation and disagreement arising from other sources of annotation variability.

\paragraph{Data-Centric Machine Learning.} Data-centric machine learning highlights that improvements in data quality can yield larger gains than increasingly complex model architectures \citep{sculley2015hidden, sambasivan2021everyone, northcutt2021pervasive}. This movement reframes dataset construction, annotation, and curation as core components of the ML development pipeline rather than as pre-processing steps. Data-centric research investigates methods to detect, characterize, and correct low-quality or mislabeled data, often combining automated auditing tools with human-in-the-loop validation \citep{northcutt2021confident, paullada2021data, geiger2021garbage}. Our work contributes to this paradigm by providing a formal statistical framework for reasoning about labeling quality and annotator reliability. Whereas most data-centric approaches focus on dataset-level diagnostics or automated error detection, we model the underlying measurement process that generates label variation. This perspective complements practical data auditing by explaining why certain instances, annotators, or contexts are systematically more error-prone, thereby linking the goals of data-centric AI to the foundational theory of measurement error \citep{eckman2024survey}.

\paragraph{Measurement and Machine Learning.}
Lastly, recent work frames machine learning as a problem of measurement, emphasizing that prediction and evaluation depend on how abstract constructs are operationalized in data. Jacobs and Wallach \citeyearpar{jacobs2021measurement} argue that many fairness and validity failures arise from mismatches between theoretical constructs, such as ``toxicity,'' ``risk,'' or ``merit,'' and the proxies used to represent them. Boeschoten et al. \citeyearpar{boeschoten2020fair} extend this reasoning, showing that fairness assessments can be misleading when target variables are error-prone proxies rather than true outcomes, and proposing latent-variable models to recover fair inferences on underlying constructs. Relatedly, Gruber et al. \citeyearpar{gruber2023sources} formalize errors in outcome variables as a distinct source of uncertainty, demonstrating that training on error-prone labels induces systematic bias even when errors are unsystematic at the individual level. Wallach et al. \citeyearpar{wallach2025position} further generalize this view, contending that the evaluation of generative AI systems is itself a social-scientific measurement problem requiring explicit definition, operationalization, and validation of what benchmarks purport to measure. While these works primarily focus on measurement at the level of system outputs and evaluation targets, our work turns this lens inward to the generation of labeled data itself. We treat human annotation as a measurement process in which each observed label reflects both a latent property of interest and the structured cognitive, contextual, and social factors shaping its expression.

\section{Conceptual Model}
\label{concept-model}

\subsection{The Measurement Error Framework} 
Across the social sciences, psychology, and biostatistics, researchers have long recognized that what we observe is rarely identical to what we want to measure. The standard solution is the \emph{measurement error model}, which treats an observed response as the sum of a ``true'' underlying value and an error term. Formally:
\begin{equation}
    Y_{it} = \mu_{i} + \epsilon_{it},
\end{equation}
where $\mu_{i}$ is the true latent value for individual $i$, $Y_{it}$ is the observed response for individual $i$ on trial $t$, and $\epsilon_{it}$ is an error term capturing random deviation. Here, “individual” refers broadly to the unit being measured, such as a person or experimental unit, depending on the application.

The inclusion of the trial index $t$ reflects the fact that, in principle, the same individual can be measured repeatedly. Each repetition (or ``trial'') provides a noisy realization of the respondent’s true value. For example, suppose a survey asks the same person twice how many hours of television they watched on average per day last week. Their true value $\mu_i$ might be $3$, but due to imperfect recall, they might report $2$ hours in one response and $4$ hours in another. These repeated observations differ, not because the true value changed, but because small random fluctuations (e.g., recall errors, rounding, or distraction) enter the response. The error term $\epsilon_{it}$ captures this trial-to-trial variability for a given respondent.  

In addition to variation within respondents across trials, there is also systematic variation \emph{between respondents}. Different individuals genuinely differ in their true values $\mu_{i}$. In the television example, one respondent’s true value might be $3$ hours while another’s might be $0$ hours. Classical measurement models recognize both sources of variation: (a) stable differences across individuals (true scores) and (b) random deviations around those scores due to error. This distinction is crucial because it allows researchers to separate signal from noise, partitioning the variance of observed responses into components attributable to systematic individual differences versus measurement error. 

This deceptively simple model has served as the backbone for a wide range of fields. In psychometrics, it underpins \emph{Classical Test Theory}, which defines the reliability of a test as the proportion of variance in observed scores attributable to the true score \citep{lord2008statistical}. In survey methodology, it provides the starting point for analyzing response error and bias in opinion measurement \citep{groves2011survey}. In epidemiology and econometrics, it motivates models of misclassification and attenuation bias in regression analysis \citep{carroll2006measurement, wooldridge2010econometric}. Despite differences in terminology across disciplines, the shared core idea is that observed data are imperfect indicators of an unobserved ground truth.

\subsection{Adapting the Framework to Machine Learning}
In supervised machine learning, however, the measurement problem differs: the ``true score'' is attached not to the annotator but to the instance being labeled. A minimal adaptation of the classical model is
\begin{equation}
    Y_{ijt} = \mu_{j} + \epsilon_{ijt},
\end{equation}
where $\mu_{j}$ is the true label for instance $j$, $Y_{ijt}$ is annotator $i$'s observed label during trial $t$, and $\epsilon_{ijt}$ is the associated error term.  

This formulation highlights that in ML practice, observed training data are rarely perfect reflections of the ground truth. Annotators may disagree with one another, make occasional mistakes, or interpret ambiguous inputs differently. From the perspective of model training, these discrepancies manifest as label noise. Much of the literature treats this noise as if it were homogeneous or purely random, but decades of research in other fields suggest that error is often structured. Some items are inherently more difficult to classify, some annotators are consistently more or less reliable, and even the same annotator may perform differently depending on their circumstances.

\subsection{Decomposing Annotation Error}
\begin{figure}[htbp]
    \centering
    \includegraphics[width=1\textwidth]{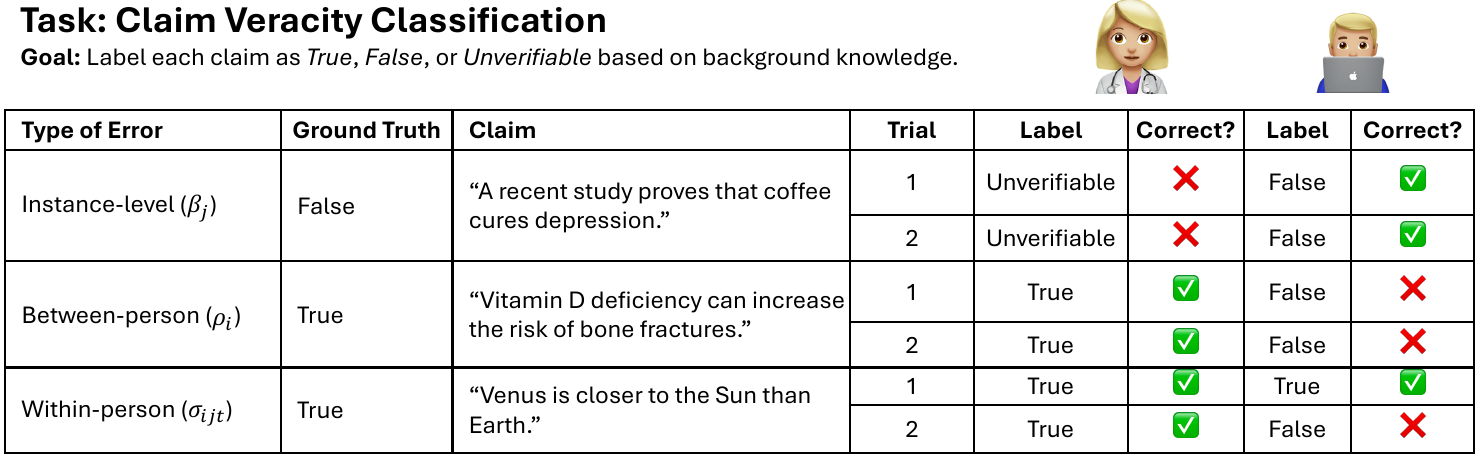}
    \caption{\textbf{Illustration of error components in the Global Ground Truth model}. Each row demonstrates a different source of annotation error using a claim verification task. \textit{Instance-level error} ($\beta_j$): Both annotators make the same mistake on an ambiguous claim where multiple interpretations are justifiable. \textit{Between-person error} ($\rho_i$): Annotators differ consistently across trials due to stable individual characteristics (here, medical expertise). \textit{Within-person error} ($\sigma_{ijt}$): The same annotator produces inconsistent labels across trials due to transient factors such as fatigue or distraction.}
    \label{fig:global-gt}
\end{figure}

Following traditions in psychometrics \citep{shavelson1992generalizability}, we move beyond the minimal model and decompose the error into three conceptually distinct components:
\begin{equation}
    \epsilon_{ijt} = \beta_{j} + \rho_{i} + \sigma_{ijt},
\end{equation}
where $\beta_{j}$ captures instance-level difficulty, $\rho_{i}$ captures stable annotator-specific tendencies, and $\sigma_{ijt}$ captures situational noise that varies across labeling sessions. Figure \ref{fig:global-gt} depicts illustrative examples of what each of these components are attempting to characterize.

\paragraph{Instance-level Error ($\beta_{j}$).}  
Instance-level error arises from attributes of the instance itself that make it inherently challenging to annotate. Even highly skilled and attentive annotators may disagree or make mistakes when the input is ambiguous or degraded. For example, in image annotation tasks, poor resolution, occlusion, or cluttered visual fields can obscure key features \citep{snow2008cheap}. In text annotation, structural or semantic ambiguity can generate uncertainty---for instance, the sentence ``There is a bird in a cage that can talk'' raises questions about whether it is the bird or the cage that possesses the ability to talk. More generally, the intrinsic difficulty of an instance often depends on its complexity: longer texts or documents with dense technical content require more cognitive effort and thus tend to produce higher error rates. Models in psychometrics and crowdsourcing frequently formalize this dimension as ``item difficulty,'' such as in Item Response Theory \citep{lord2008statistical} and probabilistic crowdsourcing models like GLAD \citep{whitehill2009whose}. In our framework, $\beta_j$ represents systematic error attributable to instance ambiguity and difficulty.  

\paragraph{Between-person Error ($\rho_{i}$).}  
Between-person error reflects persistent differences among annotators that carry across tasks and trials. These errors stem from characteristics of the annotator, such as expertise, aptitude, or personality traits. In specialized domains like medical imaging or legal document classification, domain-specific knowledge is often essential; annotators lacking sufficient training are systematically more prone to misclassification \citep{gur2003prevalence,snow2008cheap}. Beyond expertise, individual tendencies can contribute to stable differences in annotation. For example, annotators who are less conscientious may be more error-prone, while those with strong priors or ideological biases may consistently favor one label over another. In the survey methodology literature, this parallels the notion of ``response styles'' \citep{krosnick1991response}, while in crowdsourcing, it aligns with models that estimate annotator-specific accuracy and bias parameters, such as the Dawid--Skene model \citep{dawid1979maximum}. Here, $\rho_{i}$ captures those systematic annotator-specific sources of error.  

\paragraph{Within-person Error ($\sigma_{ijt}$).}  
Within-person error captures the trial-to-trial variation in annotation that arises from situational factors affecting an annotator’s performance. Even knowledgeable and conscientious annotators can produce inconsistent labels if their immediate circumstances interfere with careful judgment. For example, fatigue, distraction, or hunger can reduce attentiveness, leading to transient mistakes \citep{hauser2019common}. Similarly, environmental conditions---such as working in a noisy or uncomfortable setting---can diminish labeling accuracy. Unlike between-person error, which reflects enduring annotator characteristics, within-person error represents ephemeral influences that vary across trials and even within a single labeling session. In psychometric terms, this source of error corresponds most closely to ``state-related'' variability \citep{shavelson1992generalizability}. In practice, such error is often modeled as unsystematic noise, but acknowledging it explicitly helps distinguish between persistent annotator tendencies and situational fluctuations.

\subsection{From Conceptual to Statistical Models}

\begin{figure}[htbp]
    \centering
    \includegraphics[]{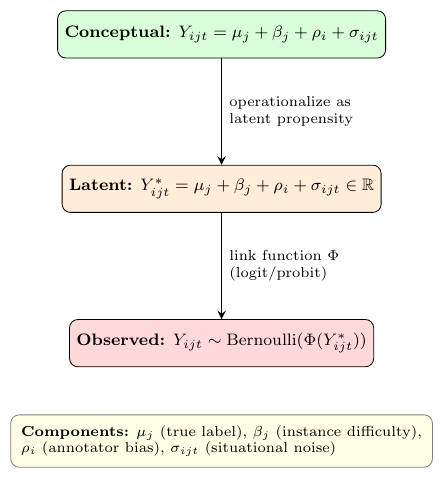}
    \caption{\textbf{Measurement model structure.} The conceptual error decomposition (top) is operationalized as a latent propensity $Y^*_{ijt}$ (middle), which is transformed via link function $\Phi$ to generate observed categorical labels (bottom). The additive structure on the latent scale implies multiplicative effects on the probability/odds scale.}
    \label{fig:measurement_model}
\end{figure}

The decomposition in Equations 2--3 describes the structure of annotation error on a latent continuous scale. For categorical labeling tasks, this latent structure must be linked to observed responses through an appropriate measurement model (Figure \ref{fig:measurement_model}).

Following the tradition of Item Response Theory \citep{lord2008statistical} and generalized linear mixed models, we posit that each annotator $i$ evaluating instance $j$ has a latent propensity $Y^*_{ijt}$ to assign the correct label:

\begin{equation}
Y^*_{ijt} = \mu_j + \beta_j + \rho_i + \sigma_{ijt}
\end{equation}

This latent propensity determines the probability of observed categorical responses through a link function. For binary outcomes (correct/incorrect):

\begin{equation}
P(Y_{ijt} = \text{correct}) = \Phi(Y^*_{ijt})
\end{equation}

where $\Phi$ is the cumulative distribution function of an appropriate distribution (logistic for logit models, normal for probit). The observed label represents a realization from this probability distribution.

Importantly, the additive structure on the latent scale implies that the effects of instance difficulty, annotator tendencies, and situational noise combine multiplicatively on the probability/odds scale. A highly difficult instance (large $|\beta_j|$) or unreliable annotator (large $|\rho_i|$) shifts the latent propensity, changing the probability of error in a nonlinear fashion.

This framework generalizes naturally to multi-class outcomes via multinomial or ordered logit models, as we demonstrate in the empirical application in Section \ref{case-study}.

\section{Extending the Framework to Human Label Variation}
\label{app:ind-truth}

The conceptual model presented above assumes a single underlying ground truth $\mu_j$ for each instance $j$. This assumption aligns with classical test theory and with the dominant paradigm in supervised machine learning, where the goal is to reproduce a unique, objective label for every observation. Recent work in NLP and ML increasingly challenges this assumption by emphasizing \emph{human label variation} (HLV)---the idea that annotators may hold multiple, equally defensible interpretations of the same instance due to differences in perspective, experience, background knowledge, or values.

Under an HLV perspective, the ``true'' label for an item is not a single fixed point but a \emph{latent interpretive construct}. Different annotators may perceive different aspects of the same instance as relevant or salient, yielding stable, systematic differences in their intended labels. Thus, rather than assuming a single latent truth $\mu_j$, we acknowledge annotator-specific latent truths $\mu_{ij}$ that govern the labels each annotator believes to be correct. This view is especially relevant for subjective or socially grounded tasks, such as toxicity, sentiment, or stance, where disagreement arises from genuine interpretive diversity rather than error.

Incorporating human label variation into the measurement framework therefore requires relaxing the assumption of a single population-level latent truth and explicitly modeling individual-level interpretations. Conceptually, we treat
$Y_{ijt}$ as a categorical observation informed by an annotator-specific latent construct $\mu_{ij}$, together with noise arising from the labeling process. Here, $\mu_{ij}$ is not a numeric label but a latent representation of annotator $i$'s intended meaning for instance $j$, and transient deviations from this intended response are captured by $\epsilon_{ijt}$.

Individual-level latent truths can themselves be decomposed into a population-level construct and annotator-specific interpretive deviations:
\[
\mu_{ij} = \mu_j + \delta_{ij},
\]
where $\mu_j$ represents the shared population-level interpretation of the item, and $\delta_{ij}$ captures stable, systematic differences in how annotator $i$ interprets or evaluates that item. The term $\delta_{ij}$ thus formalizes the idea of \emph{individual ground truth}: annotators may differ in their internal representation of what the correct label should be, even before situational noise or task difficulty is considered.

Importantly, allowing annotator-specific interpretive variation does not eliminate other sources of measurement error. Even when annotators hold stable interpretive stances, some instances are more ambiguous or complex than others, making them harder to classify and increasing disagreement regardless of who labels them. Likewise, annotators may exhibit stable tendencies across tasks (e.g., stricter vs.\ more permissive labeling styles), and momentary fluctuations such as fatigue or distraction introduce additional within-person variability.

\begin{figure}[htbp]
    \centering
    \includegraphics[width=1\textwidth]{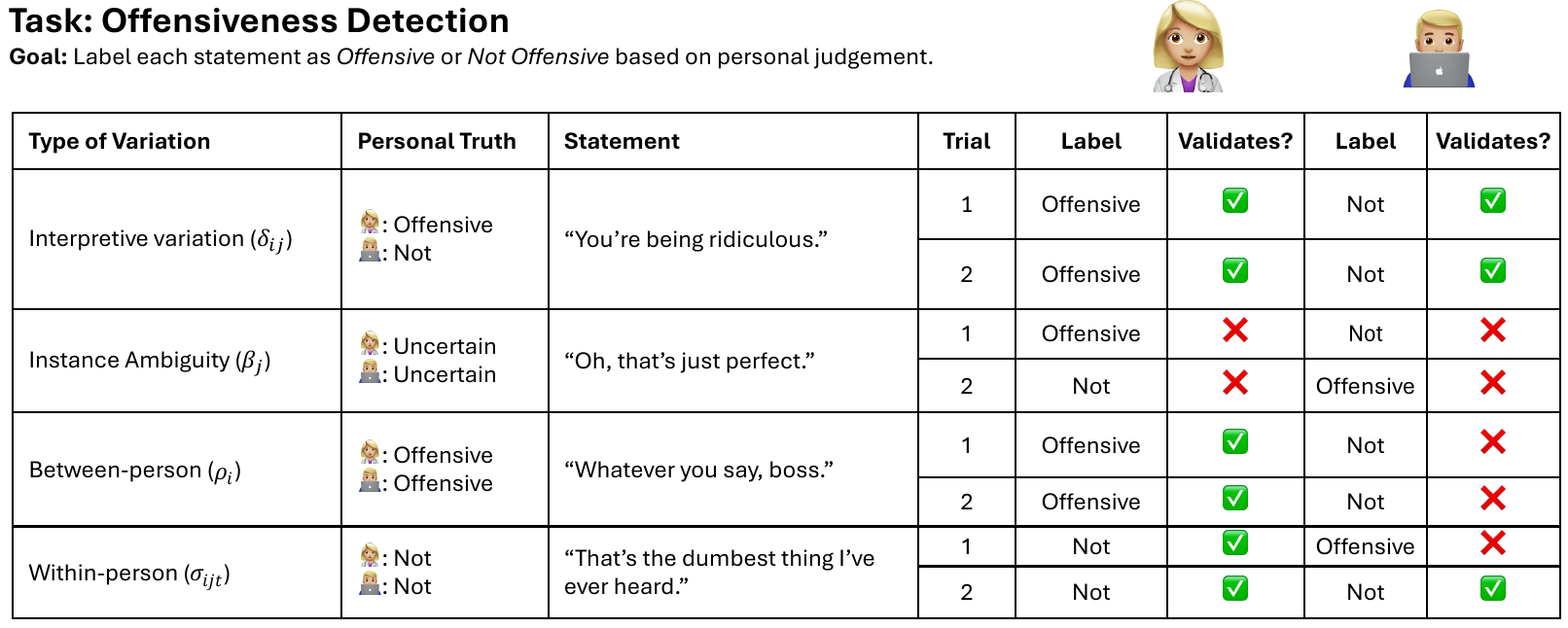}
    \caption{\textbf{Illustration of variation components in the Individual Ground Truth model.} \textit{Interpretive variation} ($\delta_{ij}$): Both annotators are internally consistent but apply different offensiveness thresholds.  \textit{Instance ambiguity} ($\beta_j$): Context-dependence creates genuine uncertainty; neither annotator can consistently self-validate any interpretation across trials, demonstrating that some instances resist stable judgment even within individual frameworks. \textit{Between-person error} ($\rho_i$): Both annotators believe sarcasm is offensive, but Annotator 2 has a stable tendency to miss subtle cues, consistently failing self-validation. \textit{Within-person error} ($\sigma_{ijt}$): Annotator 2 makes a transient error in Trial 1 but shares Annotator 1's interpretive standard. \textbf{Note:} ``Validates?'' indicates whether the annotator, upon later review, affirms their own prior judgment ($\checkmark$ = self-consistent; $\times$ = self-contradictory). Each annotator is evaluated against their own interpretive framework, not an external ground truth.}
    \label{fig:ind-gt}
\end{figure}

To accommodate these multiple sources of variation, the full conceptual model distinguishes between variation in \emph{what annotators believe the truth to be} and variation in \emph{how reliably they express that belief}. We therefore view the observed label as arising from:
\begin{equation}
Y_{ijt} \sim f\bigl(\mu_j + \delta_{ij},\; \beta_j,\; \rho_i,\; \sigma_{ijt}\bigr),
\end{equation}
where $\mu_j$ and $\delta_{ij}$ jointly describe the latent truth as perceived by annotator $i$, $\beta_j$ captures shared instance-level ambiguity or difficulty that makes it hard for any annotator, $\rho_i$ captures stable annotator-level tendencies across items, and $\sigma_{ijt}$ captures transient within-person noise. The function $f(\cdot)$ represents the process by which latent constructs and error components generate categorical observations. Figure \ref{fig:ind-gt} depicts illustrative examples of these recontextualized components.

This formulation makes explicit that disagreement can arise from at least two separable sources: (1) \emph{interpretive variation} in what different annotators believe the correct label should be, and (2) \emph{measurement error} in how consistently annotators express those beliefs. Together, these components define the structure of human label variation in annotation data and clarify when a single shared ground truth is appropriate and when interpretive diversity must be modeled directly.

\section{Choosing the Appropriate Ground Truth Perspective}

The framework above highlights that not all labeling tasks share the same underlying notion of ``truth.''  
For some domains, such as object recognition or digit classification, it may be reasonable to assume a single, shared ground truth ($\mu_j$) to which all annotators approximate. In others, such as sentiment analysis or toxicity detection, disagreement may reflect multiple coherent interpretations rather than random error. Determining which of these regimes a dataset occupies is not trivial: it depends on whether disagreement arises primarily from shared item difficulty, from systematic annotator differences, or from stable relational patterns among annotators. In measurement terms, the question is whether the construct being labeled functions as a single latent variable or as a family of context-dependent judgments. Before deploying supervised learning models, researchers must therefore assess which interpretation of ``truth'' is empirically more warranted for the task at hand.

\paragraph{Connection to Classical Measurement Theory.}
In traditional measurement research, determining whether a construct can be represented by a single latent dimension is a prerequisite to modeling it.  
Psychometricians routinely test for unidimensionality \citep{segars1997assessing} and measurement invariance \citep{putnick2016measurement}; that is, whether different indicators or raters reflect the same underlying trait in comparable ways.  
If these conditions fail, a single composite score is not meaningful, and the construct must be treated as multidimensional or context-dependent.  
The same logic applies to labeled data in machine learning: before treating consensus labels as ground truth, we must ask whether the annotation process behaves as if all annotators are measuring the same latent variable.  
When disagreement is dominated by instance-level difficulty, the global-truth model is appropriate; when disagreement is structured across annotators or relationships, the assumption of a single latent truth breaks down, motivating individualized or rater-aware approaches.  

\subsection{Operationalizing the Framework: A General Modeling Strategy}
\label{sec:operationalization}
This logic can be implemented empirically using hierarchical models that partition the variance in labeling outcomes into interpretable components.  
At a conceptual level, the approach estimates how much of the observed disagreement among annotators can be attributed to three sources:  
(a) systematic properties of the item being labeled (instance difficulty, $\beta_j$),  
(b) stable annotator-specific tendencies ($\rho_i$), and  
(c) residual or interaction-level variability ($\sigma_{ijt}$ and, where data permit, $\delta_{ij}$).  
In practice, this can be achieved through mixed-effects or generalized linear mixed models (GLMMs), which treat annotations as nested observations within items and annotators.  
These models estimate fixed effects for measurable item characteristics and random effects for annotator and item identities, thereby decomposing disagreement into structured and unstructured components.  
The same framework naturally extends to relational designs, such as pairwise validation, where additional random effects capture compatibility between annotators or between annotators and items.  
Together, these models provide a general method for testing whether disagreement in a labeling task more closely reflects a single shared construct (favoring a global-truth perspective) or structured interpretive variation (favoring an individual or HLV perspective).  

\section{Case Study: VariErr NLI}
\label{case-study}

To demonstrate the practical utility of the measurement-error framework, we apply it to the \emph{Variation versus Error Natural Language Inference (VariErr NLI)} dataset \citep{weber-genzel-etal-2024-varierr}. 
Natural language inference (NLI) requires deciding whether a ``hypothesis'' statement logically follows from a ``premise'' statement, with labels of \emph{entailment} (hypothesis follows), \emph{contradiction} (hypothesis does not follow), or \emph{neutral} (premise and hypothesis unrelated). 
VariErr draws 500 items from the MNLI subset of ChaosNLI \citep{nie-etal-2020-learn}, each labeled by four independent annotators.

The VariErr data is ideal for this work because it provides a richer annotation structure than conventional NLI datasets. 
Annotators supplied not only categorical labels but also written explanations justifying their decisions. 
These \emph{ecologically valid explanations} \citep{jiang-etal-2023-ecologically} capture reasoning at the time of labeling rather than retrospective rationalization. 
Two months later, the same annotators re-evaluated the pooled (label, explanation) pairs under blinded conditions, producing a second round of annotations that enable direct assessment of within- and between-annotator consistency. 
This multi-round design allows estimation of both instance-level difficulty and stable annotator differences, as well as the relational structure of mutual validation.

\subsection{Modeling Structured Sources of Annotation Error}

Following the modeling strategy outlined in Section~\ref{sec:operationalization}, we treat each annotation as a nested observation within annotator and item. 
We fit mixed-effects logistic regressions to estimate how instance- and annotator-level factors contribute to labeling errors, using both global and individual ground truth definitions described above.

\textbf{Global Ground Truth Model.} For the global-truth model, we operationalize error as $Z_{ijt} = \mathbbm{1}(Y_{ijt} \neq \mu_j)$, where $Y_{ijt}$ is annotator $i$'s observed categorical label for instance $j$ on trial $t$, and $\mu_j$ is the inferred consensus label. We estimate:

\begin{equation}
\text{logit } P(Z_{ijt} = 1) = \mathbf{X}_j\boldsymbol{\beta} + \rho_i + \sigma_{ijt}
\end{equation}

where $Z_{ijt} = 1$ indicates an error (deviation from consensus). The vector $X_{j}$ contains item-level predictors such as ambiguity, sentence length, and lexical overlap, and $\beta$ represents the corresponding fixed-effect coefficients. The random intercept $\rho_{i}$ captures stable annotator tendencies across items, while $\sigma_{ijt}$ represents residual trial-level variation. 

\textbf{Individual Truth Model.} For the individual-truth model, error is defined relative 
to each annotator's self-assessed interpretation. We define 
$Z^{\text{indiv}}_{ij} = \mathbbm{1}(Y_{ij} \neq \mu_{ij})$, where $\mu_{ij}$ represents 
annotator $i$'s self-validated label for instance $j$. In practice, $Z^{\text{indiv}}_{ij} = 1$ 
when annotator $i$ later rejected their own (label, explanation) pair as incoherent. 
We estimate:

\begin{equation}
\text{logit } P(Z^{\text{indiv}}_{ij} = 1) = \mathbf{X}_j\boldsymbol{\beta} + \rho_i
\end{equation}

Because each annotator evaluates the validity of their own annotations, this specification 
omits trial-level variability ($\sigma_{ijt}$), meaning residual variation conflates any 
instability in personal interpretation with situational factors during validation.

Further details on the construction of these outcome variables and instance features are provided in Appendix~\ref{app:varierr-transform}.

\subsection{Measurement Design Diagnostic}
\label{sec:diagnostic}

The previous models estimate how instance- and annotator-level factors contribute to labeling error under alternative definitions of truth.  
We now introduce a complementary diagnostic that evaluates whether a labeling task behaves as if it reflects a single shared construct or structured individual variation.  
Rather than comparing annotators to an external ground truth, this approach examines the \emph{relational structure} of agreement among annotators themselves.  
We posit that if disagreement arises primarily from shared instance difficulty, we should observe high labeler--judge consistency once item effects are accounted for.  
If, however, disagreement follows stable patterns across annotators or annotator pairs, this indicates systematic interpretive alignment, supporting an individual or human label variation (HLV) perspective.

Formally, we model the probability that judge $k$ validates labeler $i$’s annotation for item $j$ as:

\begin{equation}
    \text{logit}\big(P(V_{kij} = 1)\big)
    = \alpha 
    + u^{(l)}_{i} 
    + u^{(j)}_{k} 
    + u^{(t)}_{j} 
    + u^{(lj)}_{ik},
\end{equation}

where $V_{kij} = 1$ if judge $k$ validated labeler $i$’s label for item $j$; 
$u^{(l)}_{i}$, $u^{(j)}_{k}$, and $u^{(t)}_{j}$ are random intercepts for labeler, judge, and item, respectively; 
and $u^{(lj)}_{ik}$ is a random labeler--judge interaction capturing structured relational alignment (i.e., if some pairs of annotators understand each other better than others, even after accounting for labeler and judge effects).  
The variance of each component ($\sigma_l^2$, $\sigma_j^2$, $\sigma_t^2$, and $\sigma_{lj}^2$) quantifies how strongly that source contributes to overall variation in validation probability.

The relative magnitudes of these variance components serve as a diagnostic of the underlying \emph{measurement regime}:

\begin{itemize}
    \item \textbf{Global Ground Truth regime.}  
    Disagreement is dominated by shared instance difficulty.  
    The ideal variance profile is characterized by:
    \[
        \sigma_t^2 \gg \sigma_l^2, \; \sigma_j^2, \; \sigma_{lj}^2.
    \]
    High item-level variance ($u^{(t)}_j$) indicates that some instances are broadly confusing or ambiguous, but annotators otherwise behave similarly.  
    Producer and judge effects are small once item difficulty is considered, suggesting that all annotators approximate a single common construct.

    \item \textbf{Individual Ground Truth regime.}  
    Disagreement reflects stable individual tendencies and structured pairwise alignment.  
    The ideal variance profile reverses:
    \[
        \sigma_l^2 \text{ and } \sigma_{lj}^2 \gg \sigma_t^2.
    \]
    High labeler- or judge-level variance ($u^{(l)}_i$, $u^{(j)}_k$) indicates persistent personal bias, reliability, or interpretive stance.  
    High interaction variance ($u^{(lj)}_{ik}$) reveals consistent pairwise agreement patterns---some annotator pairs systematically validate each other’s reasoning, implying interpretive subcommunities.  
    Together, these patterns reflect a task better modeled as aggregating coherent but divergent human perspectives rather than measuring deviations from a single truth.
\end{itemize}

In practice, most labeling tasks are unlikely to conform perfectly to either extreme.  
Tasks involving complex or socially grounded judgments may empirically display a hybrid of the two regimes, where shared ambiguity and stable individual interpretation coexist. Some items may be genuinely confusing for everyone (global difficulty), while others invite divergent yet internally coherent readings (individual interpretation). Identifying the relative balance of these components therefore helps determine whether a dataset behaves more like a shared measurement of one construct or a structured aggregation of plural perspectives.

In summary, this relational diagnostic complements the earlier mixed-effects models by providing a direct, data-driven assessment of the structure of disagreement. A dominance of item-level variance supports a shared or ``global'' ground-truth interpretation, whereas dominance of labeler and pairwise variances supports an individual-truth or HLV interpretation.

\section{Results}

In the Global Ground Truth model (Table 1), we assumed a shared consensus label for each instance and modeled deviations from this as annotation error. To closely match our conceptual model, we first fit a random-effects–only model across documents, labelers, and trials to capture structured variability that comes from the data’s grouping or hierarchy.  This model reveals substantial variance across documents ($\text{SD}=0.74$), labelers ($\text{SD}=0.30$), and trials ($\text{SD}=0.22$), confirming our theory that annotation errors are structured rather than purely random. We then replace the document random effect with lexical and structural text variables to test the hypothesis that NLI errors can be explained by features of the text. None of these new fixed effects variables were statistically significant, except for the feature capturing negation flip between the premise and hypothesis statement. Introducing ambiguity as a fixed predictor (a post-hoc diagnostic derived from annotation patterns themselves; see Appendix A.2) dramatically improved model fit ($\Delta \text{AIC} \approx -560$), with ambiguous items roughly nine times more likely to be mislabeled. Other lexical or structural features remained near-null once ambiguity was included, suggesting that semantic ambiguity rather than surface textual form is the dominant source of instance-level error. Additionally, random intercepts for annotators and trials remained nonzero even in the full model, indicating persistent between-annotator and within-person variability in labeling accuracy.

\begin{table}[t]
\centering
\caption{Mixed-effects logistic regression models predicting annotation error under the global ground truth definition.}
\label{tab:model_global}
\begin{threeparttable}
\small
\begin{tabular}{lccc}
\toprule
 & \textbf{Random-only} & \textbf{Baseline Features} & \textbf{+ Ambiguity} \\
\midrule
\textbf{Fixed Effects (OR [95\% CI])} & & & \\
Intercept & 0.22 [0.14, 0.34]$^{***}$ & 0.25 [0.16, 0.37]$^{***}$ & 0.06 [0.04, 0.10]$^{***}$ \\
Ambiguity (TRUE) & --- & --- & 9.05 [7.31, 11.20]$^{***}$ \\
Lexical Overlap (1 SD) & --- & 0.99 [0.91, 1.07] & 1.02 [0.93, 1.11] \\
Avg. Toks/Sent (1 SD) & --- & 0.98 [0.90, 1.06] & 1.06 [0.96, 1.17] \\
Neg. Presence Flip & --- & 1.22 [1.02, 1.47]$^{*}$ & 0.97 [0.80, 1.18] \\
Entity Jaccard (1 SD) & --- & 0.98 [0.90, 1.06] & 1.00 [0.91, 1.09] \\
Norm. Overlap (1 SD) & --- & 1.03 [0.95, 1.12] & 1.01 [0.93, 1.11] \\
\midrule
\textbf{Random Effects (Var / SD)} & & & \\
Document (Intercept) & 0.55 / 0.74 & --- & --- \\
Labeler (Intercept) & 0.09 / 0.30 & 0.08 / 0.28 & 0.09 / 0.30 \\
Trial (Intercept) & 0.05 / 0.22 & 0.04 / 0.21 & 0.06 / 0.24 \\
\midrule
\textbf{Model Fit} & & & \\
Log-Likelihood & $-1916.55$ & $-1943.64$ & $-1663.86$ \\
AIC & $3841.1$ & $3903.3$ & $3345.7$ \\
$N$ & 3814 & 3814 & 3814 \\
\bottomrule
\end{tabular}
\begin{tablenotes}
\footnotesize
\item ORs are odds ratios with 95\% Wald confidence intervals. Continuous predictors standardized (per 1 SD). 
Significance: $^{*}p<.05$, $^{**}p<.01$, $^{***}p<.001$.
\end{tablenotes}
\end{threeparttable}
\end{table}

The Individual Ground Truth model (Table 2) reframes “error” from each annotator’s perspective, treating their own self-evaluations as the reference standard. A random-effects–only baseline showed large between-labeler variance ($\text{SD}=1.19$) and negligible document-level variance ($\text{SD}=0.05$), in stark contrast to the Global Ground Truth model with had relatively larger document-level variance and smaller between-labeler variance. This implies that, under the Individual Ground Truth assumption, disagreement primarily reflects stable annotator tendencies rather than inherent item difficulty and challenges folk wisdom that more or better data would resolve the disagreement. Strikingly, while in the Global Ground Truth model ambiguity predicts more error (shared confusion), in the Individual Ground Truth model it predicts fewer self-perceived errors (OR = 0.68 [0.50, 0.92], $p = .013$). This result suggests annotators appeared to interpret their own ambiguous decisions as internally coherent, even when others disagreed. Like in the Global Ground Truth models, the textual and lexical features do not explain much about individual annotator's self-perceived correctness, indicating that most structured variation is captured by stable annotator-specific effects rather than measurable item features.

\begin{table}[t]
\centering
\caption{Mixed-effects logistic regression models predicting annotation error under the individual ground truth definition.}
\label{tab:model_individual}
\begin{threeparttable}
\small
\begin{tabular}{lccc}
\toprule
 & \textbf{Random-only} & \textbf{Baseline Features} & \textbf{+ Ambiguity} \\
\midrule
\textbf{Fixed Effects (OR [95\% CI])} & & & \\
Intercept & 0.08 [0.02, 0.26]$^{***}$ & 0.07 [0.02, 0.24]$^{***}$ & 0.09 [0.03, 0.29]$^{***}$ \\
Ambiguity (TRUE) & --- & --- & 0.68 [0.50, 0.92]$^{*}$ \\
Lexical Overlap (1 SD) & --- & 1.03 [0.88, 1.20] & 1.02 [0.87, 1.19] \\
Avg. Toks/Sent (1 SD) & --- & 1.14 [1.01, 1.29]$^{*}$ & 1.12 [0.99, 1.28]$^{\dagger}$ \\
Neg. Presence Flip & --- & 1.14 [0.81, 1.62] & 1.21 [0.85, 1.71] \\
Entity Jaccard (1 SD) & --- & 0.98 [0.84, 1.16] & 0.98 [0.83, 1.15] \\
Norm. Overlap (1 SD) & --- & 1.17 [0.97, 1.40] & 1.17 [0.98, 1.41]$^{\dagger}$ \\
\midrule
\textbf{Random Effects (Var / SD)} & & & \\
Document (Intercept) & 0.00 / 0.05 & --- & --- \\
Labeler (Intercept) & 1.41 / 1.19 & 1.41 / 1.19 & 1.39 / 1.18 \\
\midrule
\textbf{Model Fit} & & & \\
Log-Likelihood & $-590.99$ & $-586.79$ & $-583.68$ \\
AIC & $1188.0$ & $1187.6$ & $1183.4$ \\
$N$ & 1907 & 1907 & 1907 \\
\bottomrule
\end{tabular}
\begin{tablenotes}
\footnotesize
\item ORs are odds ratios with 95\% Wald confidence intervals. Continuous predictors standardized (per 1 SD). 
Significance: $^{\dagger}p<.10$, $^{*}p<.05$, $^{**}p<.01$, $^{***}p<.001$.
\end{tablenotes}
\end{threeparttable}
\end{table}

To further assess the structure of disagreement, the Pairwise Validation model (Table 3) decomposed validation outcomes across labelers, judges, items, and their interactions. This model revealed substantial variance across judges ($\text{SD}=1.08$), items ($\text{SD}=0.97$), and labeler–judge interactions ($\text{SD}=0.48$), indicating that interpretive alignment depends jointly on who produced and who evaluated a label.

The relative magnitudes of these variance components provide a diagnostic view of where disagreement originates. The sizable item-level variance ($u^{(t)}{j}$) indicates that some instances are inherently more difficult to evaluate, consistent with a global ground-truth regime in which shared ambiguity drives labeling error. At the same time, the large judge-level variance ($u^{(j)}{k}$) reveals persistent differences in how strictly individual annotators evaluate others’ labels, pointing to stable individual tendencies or biases. Most notably, the substantial labeler–judge interaction variance ($u^{(pl)}_{ik}$) shows that certain annotator pairs consistently align or diverge in their interpretations. This pattern captures structured relational alignment, the hallmark of human label variation, demonstrating that disagreement follows predictable interpretive relationships rather than random fluctuation.

The model’s high intercept (OR = 9.57 [3.20, 28.59]) corresponds to an overall validation probability of approximately 0.90, indicating broad consensus with selective pockets of disagreement rather than pervasive unreliability. In combination, these results reveal a layered structure of variation: item-level difficulty contributes to shared uncertainty, annotator-level differences capture enduring individual stances, and interaction-level variance exposes relational dynamics of agreement and tension. When such relational variance is nontrivial, the assumption of a single shared ground truth becomes less defensible, and the data are better conceptualized under a human label variation framework—one in which multiple, internally consistent interpretations coexist within a generally reliable annotation process.

Across all models, the pattern of variance points toward the individual-ground-truth perspective as a better description of the VariErr NLI task. Disagreement is not dominated by item-level ambiguity alone as would be expected under a single shared truth but by stable, annotator-specific tendencies and consistent relational alignments among annotators. In other words, for this case study, the label disagreement reflects interpretable and systematic human variation rather than random noise around a common target. This interpretation is consistent with recent work showing that disagreement in NLI arises from systematic and interpretable sources, such as differing reasoning strategies, linguistic interpretations, or epistemic stances, rather than random annotation error \citep{gruber2024more, jiang-etal-2023-ecologically, hong2025litex}.

\begin{table}[t]
\centering
\caption{Mixed-effects logistic regression model predicting pairwise validation outcomes ($V_{kij}$).}
\label{tab:model_pairwise_validation}
\begin{threeparttable}
\small
\begin{tabular}{lc}
\toprule
\textbf{} & \textbf{Estimate} \\
\midrule
\textbf{Fixed Effect (OR [95\% CI])} & \\
Intercept & 9.57 [3.20, 28.59]$^{***}$ \\
\midrule
\textbf{Random Effects (Var / SD)} & \\
Document & 0.94 / 0.97 \\
Judge & 1.16 / 1.08 \\
Labeler  & 0.02 / 0.13 \\
Labeler--Judge (Interaction) & 0.23 / 0.48 \\
\midrule
\textbf{Model Fit} & \\
Log-Likelihood & $-2987.82$ \\
AIC & $5985.65$ \\
$N$ (observations) & 7710 \\
Groups & Items = 500; Judges = 4; Labelers = 4; L--J pairs = 16 \\
\bottomrule
\end{tabular}
\begin{tablenotes}
\footnotesize
\item ORs are odds ratios with 95\% Wald confidence intervals. 
Model estimated using maximum likelihood with a binomial link.  
Significance: $^{*}p<.05$, $^{**}p<.01$, $^{***}p<.001$.
\end{tablenotes}
\end{threeparttable}
\end{table}


\section{Discussion}
This paper reframes annotation as a measurement process and demonstrates how a decomposition of variance across instances, annotators, and relationships can help diagnose the structure of disagreement in labeled data. Using mixed-effects models under two notions of truth, global consensus and individual self-assessment, together with a relational (pairwise validation) design, we find that disagreement in the VariErr NLI dataset is not dominated by shared item difficulty alone, but by stable annotator tendencies and structured labeler–judge alignments. In the language of our framework, the empirical signal concentrates on $\rho_i$ and $\delta_{ij}$ more than on $\beta_j$, indicating that this task sits closer to the human label variation (HLV) regime than to a single shared ground truth.

This work's central contribution is not another method for post hoc denoising, but a framework and diagnostic lens for deciding how to conceptualize and model ground truth in the first place. When item-level variance dominates and fixed effects such as ambiguity strongly and consistently predict error, a global-truth assumption is defensible. When annotator-level variance dominates and relational (labeler--judge) interactions are substantial, individualized or rater-aware approaches are more appropriate. In our case study, the latter pattern prevails: ambiguity does explain shared difficulty, but residual structure sits with annotators and their relationships, and ambiguity even reduces self-perceived error under individual truth. This pattern implies that preserving, modeling, and potentially training on distributions of labels, rather than collapsing to consensus, may better reflect the construct being measured.

The VariErr case study also provides empirical support for the conceptual measurement-error framework itself.  
Each theoretically proposed component of the model manifests in the data with the expected structure and magnitude. Instance-level difficulty ($\beta_{j}$) appears through the strong effect of ambiguity in the global-truth model;  annotator bias ($\rho_{i}$) is captured by the large between-labeler variance in the individual-truth model;  and structured interpretive alignment ($\delta_{ij}$) emerges in the pairwise validation model as significant producer--judge interaction variance.  Even situational noise ($\sigma_{ijt}$) is reflected in nonzero trial-level variability. The presence of all four components in the empirical data provides direct evidence that the proposed decomposition corresponds to real, measurable phenomena possible in human annotation behavior, supporting the framework’s use as both a conceptual advance and a practical diagnostic.

In addition to its theoretical contribution, this work also has several implications for ML practice. First, dataset curation should be guided by an explicit measurement perspective. If diagnostics suggest a global-truth regime, resources are best spent on disambiguating items and improving instructions. If diagnostics indicate HLV, collecting richer rater signals (calibration tasks, repeated measures, explanation quality, rater covariates) and preserving disagreement become priorities. Appendix \ref{sec:address-error} highlights additional potential strategies for addressing sources of labeling error under each of these regimes.  Second, modeling should match the regime: consensus targets and standard losses are coherent in global-truth settings; in HLV settings, multi-annotator training, mixture-of-raters, or conditional label distributions (conditioned on rater or rater clusters) are preferable. Third, evaluation should mirror the measurement assumption: accuracy against a single label may be appropriate for global truth, but HLV calls for distributional or conditional metrics (e.g., calibration to rater-conditioned label distributions or agreement with specific rater communities). By articulating how item, annotator, and relational components map to alternative truth regimes, we provide a principled way to reason about labeling pipelines before committing to a training or evaluation strategy. These choices connect directly to data-centric ML and to recent calls to align evaluation with measurement theory.

\subsection{Limitations and Future Work}

This work has several limitations that are suggestive of areas for future research. First, our decomposition assumes that item difficulty ($\beta_j$), annotator tendencies ($\rho_i$), and relational alignment ($\delta_{ij}$ or $u^{(lj)}_{ik}$) are separable given the available design. In sparse or unbalanced settings (few annotators, few repeats, limited crossing), these components can be confounded, and some random effects may approach boundary solutions. Stronger designs tailored for the purposes of studying labeling errors (more raters, crossed rather than nested structure, repeated measures over time) should improve identifiability and the stability of variance estimates. Second, while VariErr NLI was useful for demostrating how the conceptual model could be operationalized, it only reflects a single specific task and design. Future work could perform simliar decompositions on a wider variety of tasks that vary in subjectivity, domain expertise, and stakes. Third, though we model item features (e.g., ambiguity) and include random intercepts for annotators and trials, we did not include other model components implied by our conceptual model, such as rater covariates (expertise, ideology, conscientiousness), temporal dynamics (learning, fatigue), or richer context (instructions, environment). Without these, variance attributed to $\rho_i$ may conflate multiple mechanisms. Collecting annotator metadata, time-on-task, and session context would enable mixed-effects location–scale \citep{hedeker2012modeling} or dynamic models \citep{finkelman2016prediction} that separate mean tendencies from within-person variability over time. Additionally, future work could extend the framework to address other labeling phenomena identified in the literature, such as task effects \citep{kern2023annotation}, order effects \citep{beck2024order}, and the impact of annotator demographics \citep{pei2023annotator}. Lastly, while high labeler–judge interaction variance indicates structured pairwise alignment, it does not, by itself, prove epistemic pluralism.

The presence of relational structure shows that certain annotators interpret items in systematically similar or divergent ways, but the meaning of that structure depends on context. Several alternative mechanisms can generate the same statistical signature. For example, two annotators might consistently agree because they share a heuristic, such as over-weighting lexical overlap in NLI judgments, or because they learned from the same annotation guidelines or training examples. Conversely, structured disagreement might reflect differential access to expertise rather than genuine plural truths: domain specialists could cluster in one interpretive group and novices in another. Future work that highlights the representativeness of annotator pools (e.g., \cite{eckman2025aligning}) may help address these concerns. Accordingly, the diagnostic approach presented here should be read as evidence toward, not a proof of, human label variation, with further validation and contextual inquiry required before establishing epistemic meaning.

Lastly, our empirical analysis focuses on a single NLI task, reflecting the rarity of annotation datasets with sufficient structure (repeated measures, crossed annotators, validation rounds) for complete variance decomposition. However, 
this limitation points toward both methodological opportunities and practical 
recommendations. Methodologically, simplified versions of our framework can 
be applied to any multi-annotator dataset to partition variance between 
instances and annotators, even without repeated trials. Practically, the 
framework highlights design principles for future annotation efforts: 
collecting repeated judgments, using crossed sampling designs, and recording 
annotator metadata would enable richer measurement-theoretic analysis at 
modest additional cost. We view the VariErr demonstration as proof of concept 
for a broader research program testing how variance structures differ across 
task types, domains, and annotation contexts.

We conclude with some future directions implied by our work. First, this research could be a first step toward developing a science of labeling: a cumulative enterprise with theory, a mechanistic understanding of labeling processes, and a growing evidence base of what does and does not work. Such a program would mirror mature measurement traditions in psychology and survey research, where construct validity and reliability are empirically studied, synthesized in meta-analyses, and used to guide practice. Second, future research could systematically map sources and structures of label variation across domains and task types, building a taxonomy of when individual versus global ground truths are most defensible. This taxonomy would allow the community to move from ad hoc denoising toward principled measurement design, where annotation protocols, model architectures, and evaluation metrics are jointly aligned with the underlying epistemic structure of the task. Third, integrating this framework with large-scale behavioral and linguistic data opens the door to computational social measurement, using patterns of labeling disagreement as data about human reasoning, norms, and interpretive communities themselves. In this way, the study of annotation ceases to be peripheral to machine learning and becomes a scientific lens on how humans make sense of language, meaning, and truth. With the rapid advancement of AI across society, such a human-centric approach offers a way to keep human judgment at the center of what machine learning systems learn and evaluate.

\impact{This work reframes labeling as a complex human measurement process. By decomposing labeling disagreement into item-, annotator-, trial-, and relational-level components, the proposed framework helps determine whether consensus labeling is appropriate or whether disagreement reflects meaningful diversity of perspective. Applied responsibly, such diagnostics can improve both the validity and fairness of machine learning systems by guiding dataset refinement when disagreement stems from ambiguity and preserving plural viewpoints when disagreement reflects genuine interpretive variation.
More broadly, treating labeling as a process of measurement encourages transparency about how human differences shape the data used to train and evaluate models. The framework does not resolve the ethical tension between consensus and pluralism, but it makes that tension empirically visible, enabling more principled decisions about what “truth” machine learning should learn to reproduce.}

\acks{We would like to thank the RTI International Fellows Program for their generous funding of this work.}

\vskip 0.2in
\bibliography{sample}

\appendix
\section{Adapting the VariErr Dataset for the Measurement Models}
\label{app:varierr-transform}

\subsection{Outcome Variables}
Supervised learning tasks implicitly assume that all annotators approximate a single, objective truth.  
However, in subjective or socially grounded tasks such as natural language inference or toxicity detection, this assumption may not hold.  
Different annotators can interpret the same text through distinct, yet individually consistent, reasoning processes.  
To empirically compare these perspectives, we construct parallel outcome definitions corresponding to two competing views of ground truth: 
(1) a \emph{Global Ground Truth} representing consensus across annotators, and 
(2) an \emph{Individual Ground Truth} representing each annotator’s own validated interpretation.

\subsubsection{Global Ground Truth}

Under the Global Ground Truth assumption, we treat the consensus label for each instance as the correct answer and code an annotation as an error when it differs from that label. VariErr includes both categorical labels and a subsequent round in which annotators evaluate whether each (label, explanation) pair ``makes sense.'' We leverage this second-round 
information to infer a more reliable benchmark of the true label for each instance and to assess correspondence between individual annotations and this inferred truth.

We first excluded responses labeled as ``I don't know'' (\texttt{idk}) to focus on substantive labeling decisions. For each instance $j$, we then derived a global ground truth label, denoted $\mu_j$, defined as the label that received the highest total number of valid judgments across annotators. This procedure prioritizes labels whose justifications were judged as coherent rather than relying purely on majority voting over label categories.

Next, we constructed two binary error indicators to operationalize annotation error at different levels of stringency. For annotator $i$ labeling instance $j$:

\begin{itemize}
    \item \textbf{$Z^{\text{label}}_{ij}$ (label-level error):} equals 1 if annotator $i$'s 
    categorical label $Y_{ij}$ differs from the consensus $\mu_j$, and 0 otherwise. Formally, 
    $Z^{\text{label}}_{ij} = \mathbbm{1}(Y_{ij} \neq \mu_j)$. This measure captures 
    disagreement with the inferred ground truth, irrespective of explanation quality.
    
    \item \textbf{$Z^{\text{explain}}_{ij}$ (explanation-adjusted error):} refines this 
    definition by incorporating whether the annotator's explanation was judged as making sense. 
    Specifically:
    \begin{itemize}
        \item Label matches $\mu_j$ and explanation makes sense $\rightarrow$ No error (0)
        \item Label matches $\mu_j$ but explanation does not make sense $\rightarrow$ Error (1)
        \item Label differs from $\mu_j$ and explanation makes sense $\rightarrow$ Error (1)
        \item Label differs from $\mu_j$ and explanation does not make sense $\rightarrow$ No error (0)
    \end{itemize}
\end{itemize}

Rather than treating these as conceptually distinct forms of error, we interpret $Z^{\text{label}}_{ij}$ and $Z^{\text{explain}}_{ij}$ as two \emph{trials} or repeated realizations of the same underlying measurement process. Each trial represents an independent but related assessment of whether a given annotation constitutes an error, with the second 
trial providing an explanation-adjusted evaluation that reflects a slightly different operationalization of the same construct.

We restructured the data to align with this interpretation and with the notation of the measurement model. Because each annotator contributed two such assessments relative to the consensus, we treat them as repeated measurements indexed by trial ($t \in \{1, 2\}$). Specifically, we retained only cases in which the original annotator also served as the judge to ensure consistency in how explanations were interpreted. The dataset was then reshaped from wide to long format so that each (annotator, instance) pair appears twice, once for 
each trial, yielding an outcome variable $Z_{ijt}$ indexed by $t$:

\begin{align*}
Z_{ij1} &= Z^{\text{label}}_{ij} = \mathbbm{1}(Y_{ij} \neq \mu_j) \\
Z_{ij2} &= Z^{\text{explain}}_{ij} = \text{explanation-adjusted error indicator}
\end{align*}

This design provides two observations per annotator--instance pair, enabling estimation of within- and between-person sources of variation in annotation error in a manner consistent with the repeated-measurement structure of the classical measurement error model.

\subsubsection{Individual Ground Truth}

To operationalize the Individual Ground Truth perspective, we again leverage the second round of VariErr judgments in which annotators evaluated whether their own prior (label, explanation) pairs ``made sense.'' An annotation is coded as an error if the same annotator 
later judged their own prior justification as incoherent.

For annotator $i$ and instance $j$, the individual ground truth is denoted $\mu_{ij}$, representing annotator $i$'s own validated interpretation of instance $j$. The individual error indicator is formally defined as:

\[
Z^{\text{indiv}}_{ij} = \mathbbm{1}(Y_{ij} \neq \mu_{ij})
\]
where $Y_{ij}$ is annotator $i$'s original categorical label and $\mu_{ij}$ represents the 
label that annotator $i$ validated as coherent upon reevaluation. Equivalently:
\[
Z^{\text{indiv}}_{ij} =
\begin{cases}
1, & \text{if annotator } i \text{ judged their own explanation as not making sense,} \\
0, & \text{if annotator } i \text{ affirmed their own explanation as making sense.}
\end{cases}
\]

This variable captures whether an annotator retrospectively perceives inconsistency in their own reasoning, providing a within-person measure of self-perceived labeling error. By using self-evaluations as the benchmark, we shift from modeling disagreement relative to others (as in the global model) to modeling inconsistency relative to one's own interpretive standard. 

We restricted analysis to cases in which the original annotator also served as the judge to ensure interpretive consistency. Unlike the global truth operationalization, which yields two trials per annotator--instance pair ($Z_{ij1}$ and $Z_{ij2}$), the individual truth model has no natural trial structure: each annotator provides a single retrospective validation of their own work. The resulting dataset therefore contains one observation per annotator--instance pair, $Z^{\text{indiv}}_{ij}$, indicating whether that annotator still endorsed their prior judgment upon reevaluation.

This single-observation structure has modeling implications. In Equation 8 (Section 6.1), we omit the trial-level random effect $\sigma_{ijt}$ because there is no repeated measurement to estimate within-person variability. Consequently, any instability in how 
annotators validate their own prior judgments is absorbed into the residual variance rather than being separately identified as situational noise.

\subsection{Instance-level Features}

We include a set of instance-level covariates that capture structural and semantic properties of each NLI pair, shown in Table~\ref{tab:feature_defs_simple}.  
These features serve two complementary purposes: (1) to test whether systematic properties of the text predict labeling error, and (2) to illustrate which kinds of instances contribute most to disagreement under alternative ground-truth assumptions.  
The variables encompass both objective textual characteristics (e.g., length, lexical overlap, negation) and more interpretive measures such as semantic similarity and ambiguity.  
Together, they operationalize the idea that some instances are inherently more difficult or underspecified than others, allowing us to evaluate whether observed labeling errors are driven by item-level ambiguity or annotator-level variation.

The variable \texttt{ambiguity} deserves particular comment.  
Rather than being a purely textual property, it is derived from the annotations themselves: an instance is coded as ambiguous if multiple label classes were judged as valid by at least one annotator.  
This measure therefore reflects shared uncertainty among annotators rather than a fixed linguistic feature of the text.  
Including it as a covariate does not imply that ambiguity \emph{causes} error; instead, it serves as a descriptive diagnostic indicating whether mislabeling tends to cluster on instances that elicit multiple plausible interpretations.  
In this way, \texttt{ambiguity} complements the exogenous text features by summarizing where disagreement arises collectively, providing insight into whether annotation errors stem primarily from shared item difficulty or from idiosyncratic individual tendencies.

\begin{table}[t]
\small
\setlength{\tabcolsep}{4pt}
\renewcommand{\arraystretch}{1.1}
\caption{Instance-level covariates and their hypothesized relation to labeling error.}
\label{tab:feature_defs_simple}
\begin{tabularx}{\textwidth}{lX}
\toprule
\textbf{Variable} & \textbf{Description and expected relation to error} \\
\midrule
\texttt{ambiguity} &
If an instance was had more than one class deemed as valid by the annotators.  
High ambiguity increase error likelihood. \\

\texttt{similarity} &
Semantic proximity between premise and hypothesis (semantic similarity score) using the \texttt{ms-marco-MiniLM-L6-v2} cross-encoder embeddings within SentenceBERT \citep{reimers-2019-sentence-bert}.  
Mid-range similarity indicates ambiguity and increases error likelihood;  
very high or very low similarity generally reduces disagreement. \\

\texttt{lexical\_overlap} &
Proportion of shared unigrams between premise and hypothesis.  
Medium or low overlap raises error risk because inference is required;  
very high overlap can also mislead annotators on non-entailments. \\

\texttt{avg\_toks\_per\_sent} &
Average sentence length across premise and hypothesis.  
Longer sentences impose greater cognitive load and slightly increase error rates. \\

\texttt{fk\_grade} &
Flesch–Kincaid readability grade \citep{kincaid1975derivation}.  
Harder or less readable text modestly increases the chance of labeling mistakes. \\

\texttt{neg\_presence\_flip} &
Binary indicator for negation or polarity mismatch between premise and hypothesis.  
Presence of a negation flip increases error likelihood, especially for contradictions. \\

\texttt{entity\_jaccard} &
Degree of named-entity alignment (persons, locations, dates).  
Lower overlap implies substitutions or mismatches and greater labeling difficulty. \\

\texttt{num\_norm\_overlap} &
Overlap of normalized numeric values (counts, dates, quantities).  
Higher overlap reduces error probability; mismatched numbers often trigger contradictions. \\
\bottomrule
\end{tabularx}
\end{table}

\section{Addressing Error Components in Labeling Design and Modeling}
\label{sec:address-error}

The measurement error framework not only clarifies where annotation variance arises but also suggests distinct levers for improving data quality and interpretability.  
Each error component in the model, instance-level (\(\beta_j\)), between-person (\(\rho_{i}\)), and within-person (\(\sigma_{ijt}\))---points to different design or analytical strategies.  
These interventions can occur \emph{before} labeling (through task design) or \emph{after} labeling (through modeling or quality control).  
Their interpretation also depends on whether the labeling process is conceptualized under a \emph{Global Ground Truth} regime (where deviations are noise) or an \emph{Individual / HLV} regime (where deviations may encode meaningful human diversity).

In sum, the same statistical decomposition that explains error also informs data collection and analysis.  
Under a global-truth perspective, the goal is to minimize each source of error to approximate a single latent construct.  
Under an individual-truth or HLV perspective, however, certain components, especially instance difficulty and between-person variation, may reflect legitimate pluralism rather than error.  
Recognizing this distinction reframes annotation design not simply as a process of error reduction but as one of measurement design, in which we decide which kinds of human variation to suppress, and which to measure.

\begin{table}[t]
\footnotesize 
\setlength{\tabcolsep}{4pt}
\setlength{\extrarowheight}{0.3em} 
\renewcommand{\arraystretch}{1.12} 
\caption{Strategies for addressing sources of labeling error across task design and modeling.}
\label{tab:error_mitigation}
\begin{tabularx}{\textwidth}{P{0.16\textwidth} P{0.26\textwidth} P{0.28\textwidth} P{0.26\textwidth}}
\toprule
\textbf{Error Source} & \textbf{Design-time Mitigation} & \textbf{Post-hoc Modeling / Adjustment} & \textbf{Interpretation under Global vs.\ HLV} \\
\midrule
\textbf{Instance-level} ($\beta_j$) &
Clarify task framing and guidelines \citep{parrish2023picture}; pilot ambiguous items \citep{pyatkin2023design}; collect rationales to separate semantic ambiguity from misunderstanding \citep{jiang-etal-2023-ecologically}. &
Model instance difficulty \citep{whitehill2009whose}; down-weight/flag high-variance items; use probabilistic or soft labels for ambiguous cases \citep{wu2023don}. &
\textbf{Global:} shared confusion to reduce. \newline
\textbf{HLV:} meaningful interpretive variation—retain/stratify. \\
\addlinespace[4pt]
\textbf{Between-person} ($\rho_{i}$) &
Recruit/screen for expertise \citep{beck2023quality}; calibration rounds \citep{klie2024analyzing}; if HLV is of interest, sample diverse perspectives \citep{plank2022problem}. &
Model annotator behavior \citep{yan2010modeling}; bias adjustment or clustering into interpretive communities; condition models on annotator embeddings \citep{deng2023you}. &
\textbf{Global:} unwanted annotator bias—correct/reweight. \newline
\textbf{HLV:} stable, meaningful differences across individuals/groups. \\
\addlinespace[4pt]
\textbf{Within-person} ($\sigma_{ijt}$) &
Shorter sessions/breaks \citep{pandey2022modeling}; randomize item order \citep{beck2024order}; monitor response time for fatigue/inattention \citep{chew2019smart}. &
Filter rushed/inconsistent responses \citep{northcutt2021confident}; include session-level effects; treat residual as baseline noise \citep{lin2024learning}. &
\textbf{Global:} random error to minimize. \newline
\textbf{HLV:} still noise—distinguish from systematic variation. \\
\bottomrule
\end{tabularx}
\end{table}

\end{document}